\documentclass[%
reprint,
superscriptaddress,
 amsmath,amssymb,
 aps,
prl,
longbibliography,
floatfix,
]{revtex4-1}
\usepackage{graphicx}
\usepackage{color}
\usepackage{hyperref}
\usepackage{float}
\usepackage{array, multirow}
\usepackage{amsmath}
\usepackage{newtxtext}              \usepackage{booktabs}
\usepackage{xcolor}
\usepackage{soul}
\usepackage{url}
\usepackage{qcircuit}
\usepackage[margin=0.8in]{geometry}

\hypersetup{colorlinks,
    linkcolor={blue!75!black!80!yellow},
    citecolor={blue!75!black!80!yellow},
    urlcolor={blue!75!black!80!yellow}
    }

\begin{document}



\title{Quantum Simulation of Open Quantum Systems Using a Unitary Decomposition of Operators}

\author{Anthony W. Schlimgen}
\affiliation{Department of Chemistry and The James Franck Institute, The University of Chicago, Chicago, IL 60637 USA}
\author{Kade Head-Marsden}
\affiliation{John A. Paulson School of Engineering and Applied Sciences, Harvard
University, Cambridge, MA 02138, USA}
\author{LeeAnn M. Sager}
\affiliation{Department of Chemistry and The James Franck Institute, The University of Chicago, Chicago, IL 60637 USA}
\author{Prineha Narang}
\affiliation{John A. Paulson School of Engineering and Applied Sciences, Harvard
University, Cambridge, MA 02138, USA}
\author{David A. Mazziotti}
\email{damazz@uchicago.edu}
\affiliation{Department of Chemistry and The James Franck Institute, The University of Chicago, Chicago, IL 60637 USA}

\date{Submitted June 23, 2021}

\begin{abstract}
Electron transport in realistic physical and chemical systems often involves the non-trivial exchange of energy with a large environment, requiring the definition and treatment of open quantum systems.  Because the time evolution of an open quantum system employs a non-unitary operator, the simulation of open quantum systems presents a challenge for universal quantum computers constructed from only unitary operators or gates.  Here we present a general algorithm for implementing the action of any non-unitary operator on an arbitrary state on a quantum device.  We show that any quantum operator can be exactly decomposed as a linear combination of at most four unitary operators.  We demonstrate this method on a two-level system in both zero and finite temperature amplitude damping channels. The results are in agreement with classical calculations, showing promise in simulating non-unitary operations on intermediate-term and future quantum devices.
\end{abstract}

\maketitle

\noindent \textit{Introduction.\textemdash} The time evolution of an electronic state is critical to predicting many important physical phenomena including exciton transport~\cite{Popp2021}, molecular conductivity~\cite{Li2005}, chemical catalysis~\cite{Prier2013}, quantum phase transitions~\cite{Dukelsky2008,Santos2015}, and magnetization~\cite{Breuer2002}.  Often these processes involve non-trivial interactions with a large environment, requiring the definition and treatment of open quantum systems~\cite{HeadMarsden2019JCP, HeadMarsden2019PRA, Gorini1976, Montoya-Castillo2015, Reimer2019, Yoshioka2019, Nagy2019, Hartmann2019, Vicentini2019}.  Recent research has considered the simulation of open quantum systems on quantum devices~\cite{Sweke2015, Hu2020, Garcia-Perez2020, HeadMarsden2021, Motta2020}.  While the complexity of these systems makes them ideal for quantum simulation, their dependence on non-unitary time propagation presents a challenge for quantum computers whose fundamental operations (gates) are unitary~\cite{Sweke2015, Hu2020}. A variety of possible approaches to non-unitary time evolution are emerging~\cite{Wei2016, Garcia-Perez2020, Berry2015, Childs2012}, including dilation methods~\cite{Sweke2015, Sweke2016, Hu2020, HeadMarsden2021, Hu2021}, imaginary time evolution~\cite{McArdle2019, kamakari2021digital, Motta2020}, time-dependent variational methods~\cite{Endo2020}, and duality quantum algorithms~\cite{Zheng2021}. Some of these methods, however, require the factorization of the non-unitary matrix~\cite{Sweke2015, Sweke2016, Hu2020, HeadMarsden2021, Hu2021} or the solution of a potentially large system of linear equations with assumptions of locality~\cite{McArdle2019, kamakari2021digital, Motta2020}.

In this \emph{Letter} we develop and implement a general quantum algorithm to simulate non-unitary time evolution on a quantum computer in which we decompose any quantum operator into the linear combination of at most four unitary operators.  The decomposition is exact in the limit that a small parameter $\epsilon$ approaches zero.  We show that this $\epsilon$-limit can be efficiently evaluated by Richardson's extrapolation~\cite{Richardson1927}.  We demonstrate this algorithm on a two-level system in both zero and finite temperature amplitude damping channels. This formalism provides a completely general approach for simulating both unitary and non-unitary evolution of quantum systems. In contrast to other linear combination of unitaries approaches~\cite{Childs2012, Zheng2021,Berry2014a}, our algorithm uses strictly unitary operators to represent the action of a non-unitary operator on a quantum state, and then performs quantum addition of known, prepared states. Furthermore, the formalism can be used in other contexts beyond evolution of quantum dynamics, such as in Hamiltonian simulation in quantum chemical applications, where the operators of interest are also often non-unitary. \\


\noindent \textit{Theory.\textemdash}
The time evolution of an open quantum system can be performed using the Kraus formalism in the operator sum form,
\begin{equation}
    \label{eq:rho}
    \rho(t) = \sum_i M_i\rho M_i^{\dagger},
\end{equation}
where the $M_i$ are Kraus maps corresponding to different environmental channels and $\rho(t)$ is the system density matrix~\cite{Kraus1983, Breuer2002}. In general, the density matrix of a quantum system can be written as the outer product of the wavefunction with itself,
\begin{equation}
    \rho(t) = \lvert \psi(t) \rangle \langle \psi(t) \rvert.
\end{equation}
Any operator $M$ can be decomposed into a Hermitian and anti-Hermitian component,
\begin{equation}
    S = \frac{1}{2}(M + M^\dagger)
\end{equation}
and
\begin{equation}
    A = \frac{1}{2}(M - M^\dagger),
\end{equation}
such that $M = S + A$.

The Hermitian and anti-Hermitian components can be written as the sum of two exponential unitary operators each,
\begin{equation}
    S = \lim_{\epsilon \rightarrow 0} \frac{i}{2\epsilon} (e^{-i\epsilon S} - e^{i\epsilon S}),
\label{eq:symmetric_expansion}
\end{equation}
\begin{equation}
    A = \lim_{\epsilon \rightarrow 0} \frac{1}{2\epsilon} (e^{\epsilon A} - e^{-\epsilon A}),
\label{eq:antisymmetric_expansion}
\end{equation}
where $\epsilon$ is the expansion parameter. With this decomposition we can write the action of $M$ on $\lvert \psi(t) \rangle$ as a sum of at most four unitary operators, regardless of whether $M$ itself is unitary. In this fashion, any non-unitary operator can be implemented on a quantum device as a sum of unitary operators. Importantly, we note that the decomposition results in strictly unitary operators, and that the only approximation arises in the choice of the expansion parameter $\epsilon$ which does not effect the unitarity of the implemented operators.

The $\epsilon$-limit can be accelerated by Richardson's deferred approach to the limit, also known as Richardson's extrapolation~\cite{Richardson1927}.  While a small $\epsilon$ is needed to obtain accurate $S$ and $A$ matrices, a sufficiently large $\epsilon$ is needed to resolve these matrices from the noise.  Because the expansions of both $S$ and $A$ in $\epsilon$ in Eqs.~(\ref{eq:symmetric_expansion}) and~(\ref{eq:antisymmetric_expansion}) are even with their largest errors on the order of $O(\epsilon^{2})$, the density matrix in Eq.~(\ref{eq:rho}), constructed from $M=S+A$, has an even expansion in $\epsilon$ with a similar error.  Higher-order expansions of the density matrix can be generated by Richardson's extrapolation.  The extrapolation employs a deferred approach to the limit where the density matrices at two different values of $\epsilon$ are used to obtain a better approximation to the $\epsilon \rightarrow 0$ limit.  Formally, we have the following expression for the extrapolated density matrix
\begin{equation}
\rho(0) \approx \frac{\rho(\epsilon_{1}) - \rho(\epsilon_{2}) r^{2}}{1-r^{2}}
\end{equation}
where $\epsilon_{1} > \epsilon_{2}$ and the ratio $r$ is equal to $\epsilon_{1}/\epsilon_{2}$.  Unlike the input density matrices, the extrapolated $\rho(0)$ is accurate to $O(\epsilon^{4})$.  Higher-order density matrices can be obtained by iterating Richardson's extrapolation with $r^{2}$ replaced by $r^{n}$ where the integer $n$ denotes the order of the error of the input density matrices. \\

\noindent \textit{Methods.\textemdash} 
In order to implement the decomposition $M=S+A$ on a quantum computer, we utilize the following quantum state preparation,

\begin{equation}
\lvert\widetilde{\psi}\rangle = R \cdot U |\psi \rangle,
\label{eq:propagate}
\end{equation}
for,
\begin{equation}
U =
\begin{pmatrix}
S_m & 0 & 0 & 0 \\
0 & -S_p & 0 & 0 \\
0 & 0 & -A_m & 0 \\
0 & 0 & 0 & A_p \\
\end{pmatrix},
\end{equation}
where
\begin{equation}
    \begin{aligned}
    &S_m = -S_p^\dagger = ie^{-i\epsilon S} \\
    &A_m= A_p^\dagger = e^{- \epsilon A}.
    \end{aligned}
    \label{eq:4ops}
\end{equation}
Each $S$ and $A$ operator is size $2^n$, where $n$ is the number of qubits, thus $U$ is size $4\textrm{x}2^n$. The operator $U$ propogates the wavefunction, while the rotation matrix $R$ adds the components of the prepared states in the appropriate manner,
\begin{equation}
    R = \frac{1}{2}
    \begin{pmatrix}
    r & -r \\
    r & \;\; r
    \end{pmatrix},
\end{equation}
for,
\begin{equation}
    r =
    \begin{pmatrix}
    I & -I \\
    I & \;\; I \\
    \end{pmatrix}.
\end{equation}
Finally, the populations are rescalled by $\epsilon$ classically after measurement on the quantum device. Since $\epsilon$ is generally less than 1, our approach requires a relatively large number of samples, as discussed below.

\begin{figure*}[ht!]
    \centering
    \includegraphics[width = 1.\textwidth]{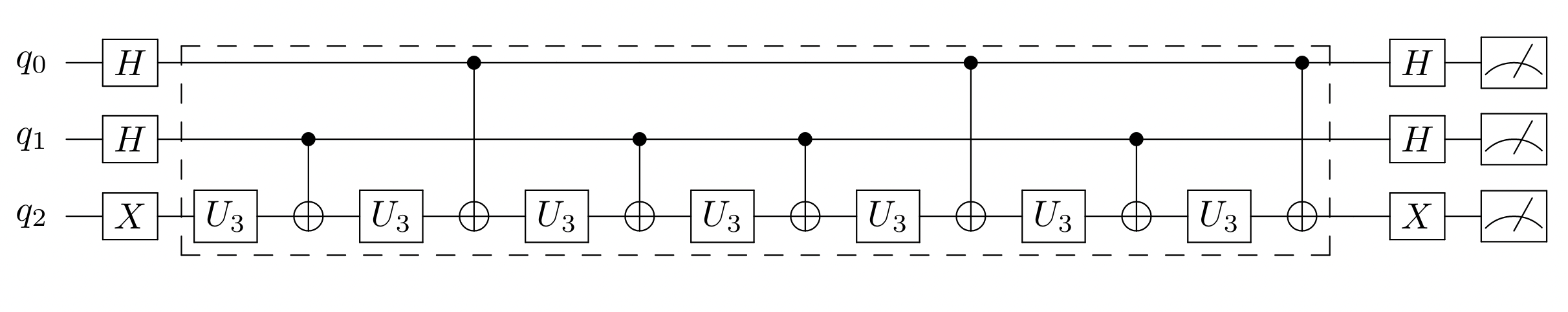}
    \caption{Circuit for preparing the wavefunction in Eq.  \ref{eq:propagate}, where the first gates represent the initial state preparation (in this case, the mixed state), the block diagonal propagation is highlighted in the dotted box, and the quantum adder is represented by the final three gates. $X$ is the X-gate, $H$ is the Hadamard gate, $U_3$ is IBM's $U_3$ gate, the 2-qubit gates are CNOT gates, and the final gates on each qubit are measurements.}
    \label{fig:circuit}
\end{figure*}

Since $U$ is diagonal, we can implement this operation using a quantum multiplexor, or uniformly controlled gate \cite{Khan2005, Roy2012,mottonen2005,Iten2016, Bergholm2005, Bullock2004, Mottonen2005a}. We utilize the uniformly controlled gate in Qiskit to propogate our initial wavefunctions~\cite{Shende2006, Qiskit}. We implement the quantum addition for the $n=3$ case with an $X$ gate on one qubit, and Hadamard gates on each of the other two qubits of the circuit. An example of this implementation is shown in Figure ~\ref{fig:circuit}.


\begin{figure}[ht!]
    \centering
    \includegraphics[scale=0.2]{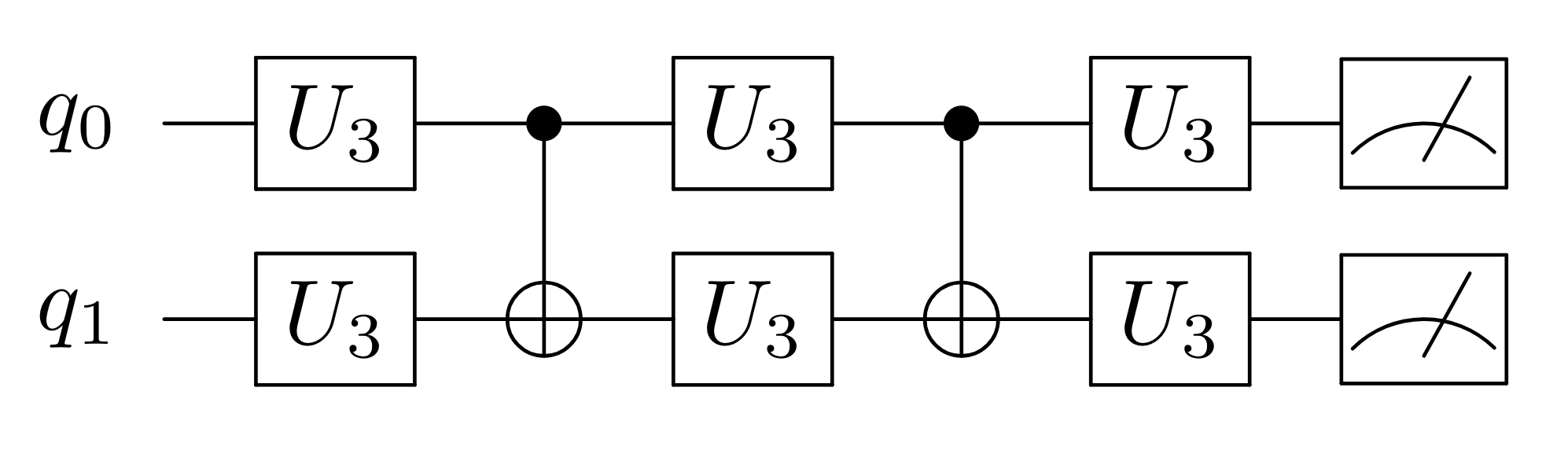}
    \caption{Example circuit for preparing a wavefunction from a $2\textrm{x}2^n$ subblock of $U$, utilizing uniformly controlled gates, or multiplexing. The state preparation and the quantum addition are included in the $U_3$ gates.}
    \label{fig:circ2}
\end{figure}

We can lower the qubit complexity of the algorithm by noting that $U$ can be implemented in parallel as 6 operators of size $2 \textrm{x} 2^n$, which are block diagonal pairs of the $S$ and $A$ operators in Eq. ~(\ref{eq:4ops}). In this approach, each $S$ and $A$ operator of size $2^n$ is also implemented resulting in an additional 4 circuits. The measured matrix elements are subsequently summed classically to reconstruct the appropriate density matrix elements. If an operator is either purely Hermitian or anti-Hermitian, only one circuit is required. We use this reduced complexity implementation to generate the data shown here. An example circuit for a $2\textrm{x}2^n$ operator is shown in Fig. ~\ref{fig:circ2}. While the three-qubit state preparation yields promising results for no-error and low-error simulations (see Supplemental Information), we observed significant drift using these circuits on the real, NISQ devices employed. As such, this paper utilizes the two-qubit preparation suited for these near-term, noisy devices.

The use of a small $\epsilon$ in the expansions in Eqs.~(\ref{eq:symmetric_expansion}) and (\ref{eq:antisymmetric_expansion}) requires a relatively large number of shots for accurate statistics on a quantum device. Our algorithm directly measures the diagonal elements of the propagated density matrix, scaled by $\epsilon$. Since the expansion in Eqs.~(\ref{eq:symmetric_expansion}) and (\ref{eq:antisymmetric_expansion}) generally requires $\epsilon<1$, and the populations are normalized to 1, these matrix elements are generally numerically small. In order to acheive good statistics in the population dynamics, many samples are therefore required on noisy devices. This drawback can be overcome by either repetition of the algorithm on shot-limited devices, or by using systems with larger available shot counts.

An alternative solution to the issue of large shot counts is the use of a Richardson extrapolation based on simulations using larger $\epsilon$ as discussed above. We use a two-point Richardson extrapolation to generate populations dynamics on quantum device, as well as using a simulator. The extrapolation allows us to achieve accurate dynamics with significantly lower shot counts using $\epsilon$'s on the order of unity ($\epsilon=1.15,1.00$), compared to simulations with smaller $\epsilon$.

\noindent \textit{Results.\textemdash} To benchmark this method, we consider a two-level system in a general amplitude damping channel. The Kraus operators are given by,
\begin{equation}
\begin{aligned}
    M_0 &= \sqrt{\lambda} \begin{pmatrix}
            1 & 0\\
            0 & \sqrt{e^{-\gamma t}}
          \end{pmatrix}\\
    M_1 &= \sqrt{\lambda} \begin{pmatrix}
            0 & \sqrt{1-e^{-\gamma t}}\\
            0 & 0
          \end{pmatrix}\\
    M_2 &= \sqrt{1-\lambda}\begin{pmatrix}
            \sqrt{e^{-\gamma t}} & 0\\
            0 & 1
          \end{pmatrix}\\
    M_3 &= \sqrt{1-\lambda}\begin{pmatrix}
            0 & 0\\
            \sqrt{1-e^{-\gamma t}} & 0
          \end{pmatrix}
          \end{aligned}
          \label{eq:gen_amp}
\end{equation}
where $\gamma$ is the decay rate, and $\lambda = \frac{1}{1+e^{-\frac{1}{K_BT}}}$ accounts for temperature dependence of the equilibrium state~\cite{Fujiwara2004, Rost2020}. In the zero temperature limit, $\lambda = 1$ and the Kraus operator evolution reduces to requiring only $M_0$ and $M_1$. For the finite temperature case, infinite temperature proves to be a good approximation for room temperature, where $\lambda = 0.5$~\cite{Fujiwara2004, Rost2020}. In this case, all four of the Kraus operators in Eq.~\ref{eq:gen_amp} are required to capture the accurate time evolution.

The initial density matrix is of the form,
\begin{equation}
    \rho(0) = \frac{1}{4}\begin{pmatrix}
            1 & 1 \\
            1 & 3\\
            \end{pmatrix},
\end{equation}
and the decay rate $\gamma = 1.52\times 10^9s^{-1}$. We chose this density matrix because of its decomposition into basis vectors which are easy to implement in a quantum circuit \cite{Hu2020}; however, this need not be the initial density matrix.

\begin{figure}[ht]
    \centering
    \includegraphics[trim=0pt 0 0 0, clip, width=0.5\textwidth]{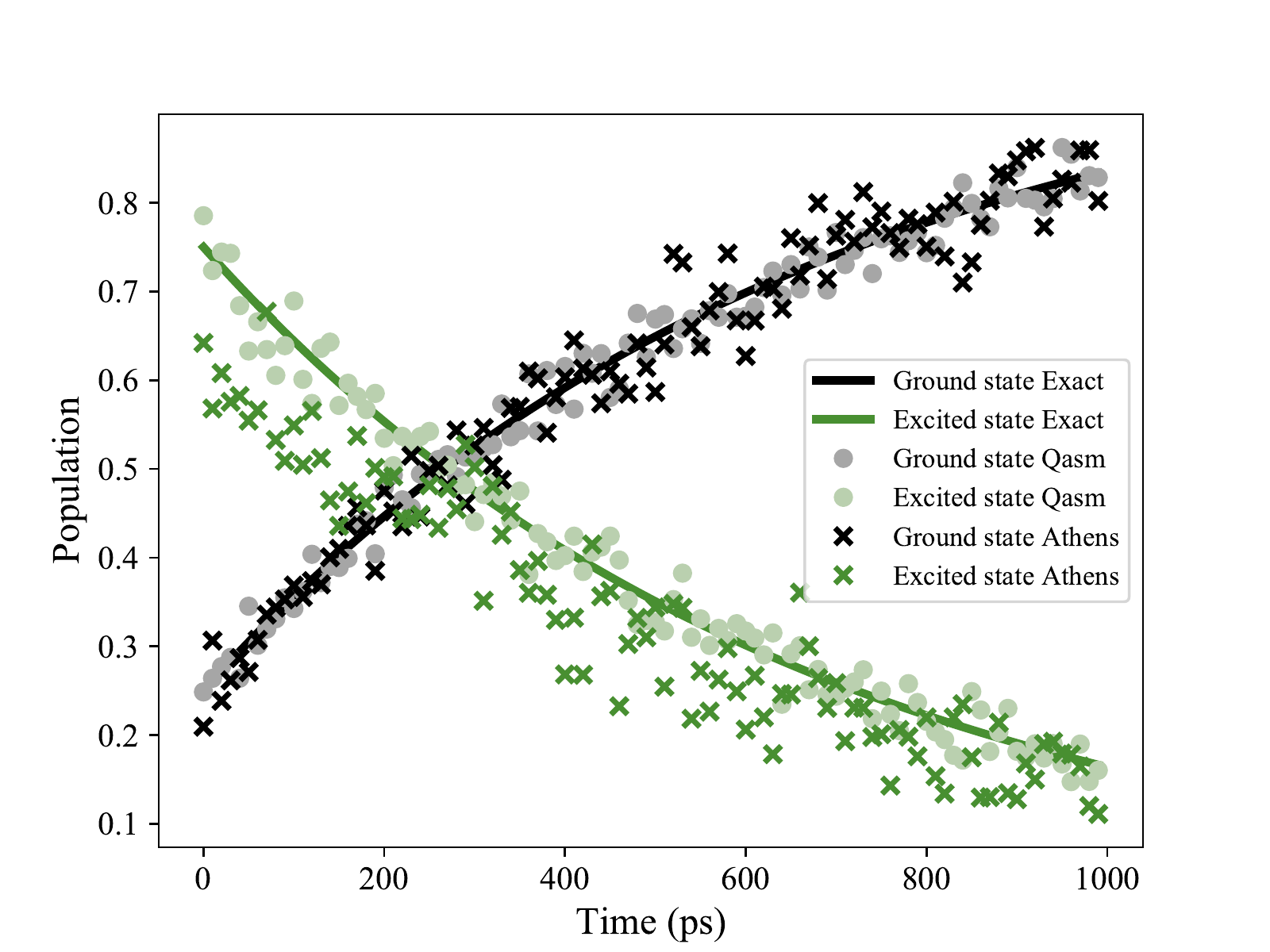}
    \caption{Excited (green) and ground (black) state populations of a two-level system in a zero temperature amplitude damping channel where lines represent the exact classical Kraus evolution, dots represent the simulation results from Qiskit Qasm simulator, and x's are results from the ibmq\_athens hardware. Qasm simulator data generated with $2^{19}$ shots and $\epsilon=0.2$. Athens results generated from the average of Richardson extrapolations ($\epsilon=1.15,1.00$) with 10 runs of $2^{13}$ shots each. The results are unmitigated with respect to error.}
    \label{fig:zero_temp}
\end{figure}

The dynamics in the zero temperature case are shown in Figure~\ref{fig:zero_temp}. The solid lines are the exact classical Kraus solution, and the dots are generated from the Qasm simulator using $2^{19}$ shots or samples~\cite{Qiskit}. We generate the simulated dynamics using the expansion presented here with $\epsilon=0.2$. Using this value of $\epsilon$, the mean absolute error between the decomposition computed classically and the exact populations is about $10^{-3}$ for the cases studied here. Finally, the x's are results from ibmq\_athens with a Richardson extrapolation~\cite{athens}. We averaged the Richardson extrapolations for 10 repetitions of $2^{13}$ shots with $\epsilon=1.15,1.00$.

\noindent Both the Qasm results using the small $\epsilon$ and the data generated from the device using the Richardson extrapolation are in good agreement with the exact solution.

\begin{figure}[ht]
    \centering
    \includegraphics[trim=5pt 0 0 0, clip,width = 0.5\textwidth]{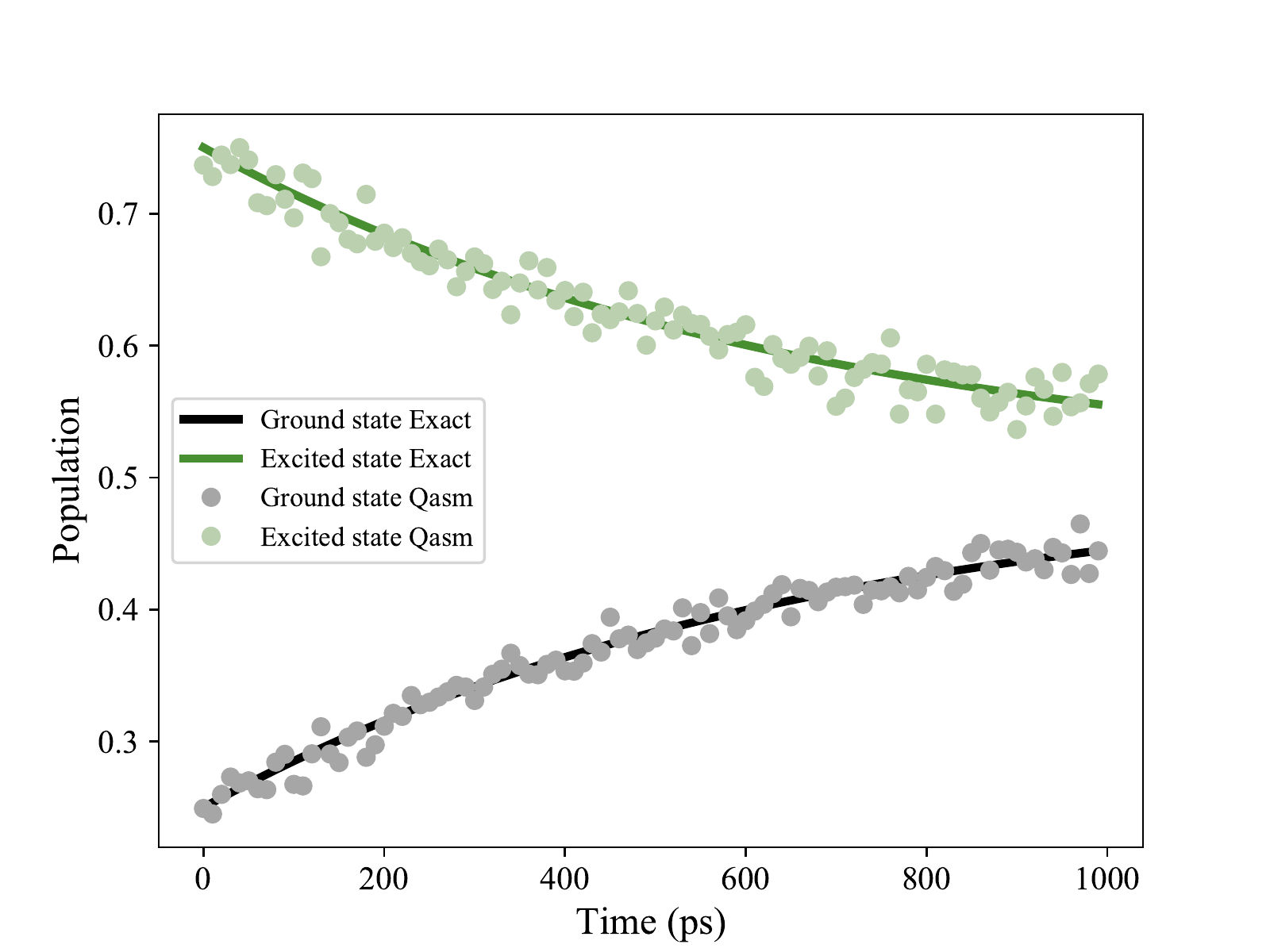}
    \caption{Excited (green) and ground (black) state populations of a two-level system in a general amplitude damping channel at room temperature where lines represent the exact classical Kraus evolution and dots represent the simulation results from Qiskit Qasm simulator using a Richardson extrapolation ($\epsilon=1.15,1.00)$, with $2^{19}$ shots }
    \label{fig:finite_temp}
\end{figure}

Initializing the system in the same initial state gives the finite temperature dynamics shown in Figure~\ref{fig:finite_temp}, where the classical Kraus evolution is again shown by the solid lines. The evolution using the Qasm simulator, $2^{19}$ shots, with a Richardson extrapolation ($\epsilon=1.15,1.00$) is shown by the dots. The simulated data show excellent agreement with the exact solution.
\\

\noindent \textit{Conclusions and Outlook.\textemdash} Simulating open quantum systems with quantum hardware presents a challenge by requiring the implementation of non-unitary processes using only unitary gates. We present a decomposition that requires the implementation of a series of operators that are strictly unitary. Our decomposition is generally applicable for any operator, either unitary or non-unitary, and can be implemented as a block-diagonal operator. The numerical approximation in the expansion parameter $\epsilon$ is the only approximation in the decomposition, but this does not affect the unitarity of the resulting (anti-)Hermitian operators.  Furthermore, we show that the convergence with $\epsilon$ can be accelerated with Richardson's extrapolation which aids the practical use of the algorithm on current devices.

This method has general applicability to problems in quantum chemistry and physics in the realm of quantum computing beyond applications in open quantum system dynamics. For example, Hermitian operators such as Hamiltonians or dipole operators, which are common in quantum chemistry, and generally non-unitary, can be implemented with the decomposition presented here. In these cases, the approach can be simplified by implementing only the Hermitian operators in Eq.~ (\ref{eq:symmetric_expansion}). Our approach would avoid Trotterization of these operators and presents an alternative that implements operators in an exponentiated unitary form which is asymptotically convergent. \\


This work is supported by the NSF RAISE-QAC-QSA, Grant No. DMR-2037783. D.A.M. also acknowledges the Department of Energy, Office of Basic Energy Sciences Grant DE-SC0019215.  We acknowledge the use of IBM Quantum services for this work. The views expressed are those of the authors, and do not reflect the official policy or position of IBM or the IBM Quantum team.


\bibliography{main}

\end{document}